\documentclass[conference,a4paper]{IEEEtran}
\usepackage{times,amsmath,epsfig,amsfonts,amssymb,graphicx,url,cite,subfigure,colortbl,bm,upgreek,multirow,booktabs}
\renewcommand{\j}{\mathrm{j}}
\renewcommand{\d}{\mathrm{d}}

\renewcommand{\baselinestretch}{1}

\DeclareRobustCommand*{\IEEEauthorrefmark}[1]{\raisebox{0pt}[0pt][0pt]{\textsuperscript{\footnotesize #1}}}
\begin{document}
\title{Total Array Gains of Millimeter-Wave \\ Mobile Phone Antennas Under Practical Conditions}
\author{\IEEEauthorblockN{
Katsuyuki Haneda,~%\IEEEauthorrefmark{1} %
Mikko Heino and %\IEEEauthorrefmark{1} and  %
Jan J\"{a}rvel\"{a}inen%\IEEEauthorrefmark{1}   %
}                                     % ...
%\\
%\IEEEauthorblockA{\IEEEauthorrefmark{1}% 1st affiliations
Department of Electronics and Nanoengineering, \\ Aalto University School of Electrical Engineering, Espoo, Finland \\
E-mail: see http://ele.aalto.fi/en/contact/}
%\IEEEauthorblockA{\IEEEauthorrefmark{3}% 2nd affiliations
%Espoo, Finland}
%}
\maketitle
\begin{abstract}
This paper studies a gain of an antenna array embedded on a mobile device operating at a millimeter-wave radio frequency. Assuming that mobile phones at millimeter-wave range operate with a single baseband unit and analog beamforming like phased arrays, we define a {\it total array gain} denoting a path gain of the phased antenna array in excess to the omni-directional path gain. The total array gain circumvents the ambiguity of conventional array gain which cannot be uniquely defined as there are multiple choices of a reference single-element antenna in an array. Two types of $8$-element patch antenna arrays implemented on a mobile phone chassis, i.e., uniform linear array (ULA) and distributed array (DA) both operating at $60$~GHz, are studied. The gain evaluated in a small-cell scenario in an airport shows that DA achieves higher median and outage gain by up to $8$ and $6$~dB than ULA when different orientations of the mobile phone are considered along with body torso and finger shadowing. There are always postures of the mobile phone where ULA cannot see the line-of-sight due to directionality of the patch antenna and of body and finger shadowing, leading to outage gain of $-15$~dB in the worst case. The DA has much smaller variation of the gain across different orientations of the phone, even when the human torso shadowing and user's finger effects are considered. 
\end{abstract}
{\smallskip \keywords Millimeter-wave, antenna array, mobile phone, array gain, user effect, body shadowing.}
\IEEEpeerreviewmaketitle

\section{Introduction}
New generations of cellular mobile networks, called the fifth-generation radios, have been studied intensively both in the industry and academy. 
The new generation of networks exploits radio frequencies higher than 6~GHz actively for higher peak data rates and network throughput.
One of the practical issues in cellular mobile radios operating at higher frequency bands is its coverage.
Higher radio frequencies typically have smaller service coverage compared to lower frequency radio, particularly in non-line-of-sight conditions in light of greater signal losses in diffraction and penetration of radio wave propagation.
Mathematical models and tools to predict losses due to radio propagation have been studied in the previous years extensively, e.g.,~\cite{Haneda16_VTCSpring, TR38901}.
In contrast, possible gains or losses attributed to antennas implemented on mobile phone devices have received less attention, particularly under the presence of fingers and human bodies of mobile users~\cite{Hejselbaek17_TAP, Syrytsin17_TAP, Zhao17_AWPL, Zhao17_TAP}.
This paper therefore sheds lights on the gains of practical mobile phone antennas operating at higher frequencies than 6 GHz, particularly at millimeter-waves (mm-waves). 

%For such frequencies, physically small mobile phone chassis becomes electrically large, allowing for implementating multiple antenna elements on the chassis. 
%One of the challenges of high frequency spatial signal processing using antenna arrays is their affordability of beamforming with fully digital implementation. 
%It will be excessive in cost and power dissipation if all the antenna elements are connected to transceiver chains through frequency up-, down-, and analog-to-digital-converters. 
%Hybrid of analog and digital beamforming is therefore discussed to reduce the number of transceiver chains~\cite{Molisch17_CM}.
%Typically the number of antenna elements supported by a single radio frequency chain is $8$ or larger, making it practical to consider mm-wave antenna arrays at a mobile phone operating like a phased antenna array~\cite{Hejselbaek17_TAP, Syrytsin17_TAP, Zhao17_AWPL, Zhao17_TAP, Jo17_TAP}.

In particular, we study an achievable gain of 60 GHz antenna arrays at a mobile phone that works as a phased antenna with a single transceiver chain.
The study considers realistic operational conditions with influence of finger, body and multipath channels. 
A novel {\it total array gain} is studied in this article, which define a received power at a mobile array that it can receive from multipath radio channels in excess to a case when a mobile is equipped with an idealistic omni-directional antenna. The definition is analogous to the mean effective gain for a single-element antenna~\cite{Vaughan87_TVT, Taga90_TVT}, but the present paper studies gains of an array.
Pathloss of a radio channel is called omni-directional pathloss when a base station (BS) and mobile station (MS) is equipped with omni-directional antennas.
The omni-directional pathloss has been studied extensively in mm-wave channel modeling, including the recently established 3GPP standard channel model for new radios~\cite{TR38901}. 
The total array gain along with the omni-pathloss therefore provides the pathloss when a mobile is equipped with an antenna array.
In contrast to the conventional array gain which cannot be defined uniquely depending on the choice of a reference single-element antenna in an array, the total array gain is uniquely defined and allows us to compare arrays formed by different antenna elements.
%In this paper, we consider the maximum ratio combining (MRC) at the mobile for a single-user downlink as it provides the maximum possible gain of the link.
We evaluate the total array gain of two $60$~GHz antenna arrays at a mobile: an $8$-element uniform linear array (ULA) and distributed array (DA) both consisting of patch antennas.
The evaluation is based on electromagnetic field simulations for antennas and measurement-based ray-tracing propagation simulations in a small-cell scenario at an airport.
Finger and human torso effects on radiation characteristics of the considered mobile phone antennas are taken into account.

The rest of the paper is organized as follows: Section~II introduces the two antenna arrays on a mobile phone and their varying postures. Effects of a finger is briefly addressed. Section~III describes ray-tracing supported by experiments, along with the body shadowing. Section~IV first derives the total array gain and compare it for the two considered antenna arrays. Section~V summarizes the main conclusions.

\section{60 GHz Mobile Phone Antennas}
\subsection{Antenna Arrays}
\begin{figure}[t]
 \begin{center}
   \subfigure[]{\includegraphics[scale=0.16]{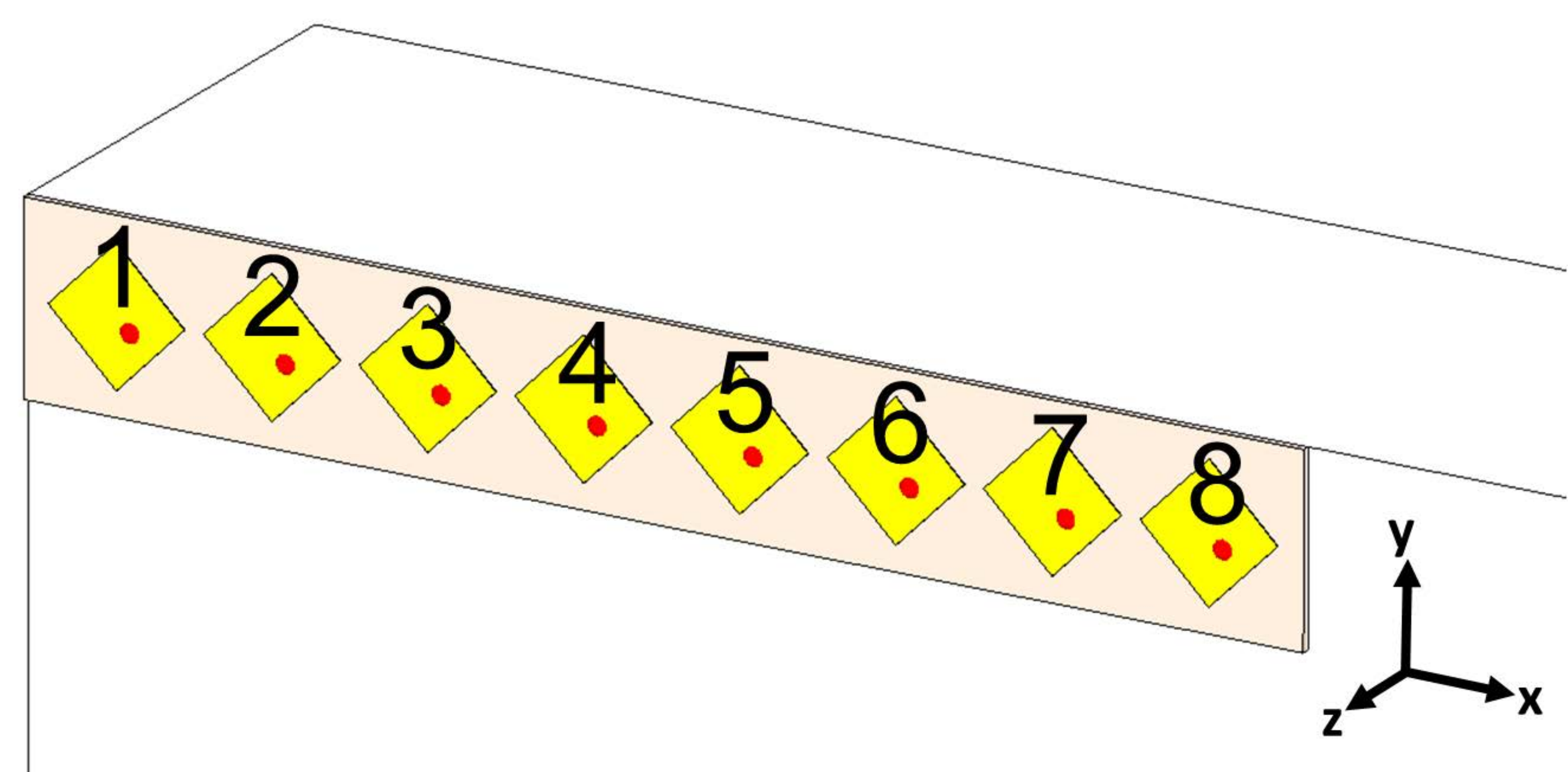}
        \label{fig:antenna_configuration_ULA}}\\
   \subfigure[]{\includegraphics[scale=0.13]{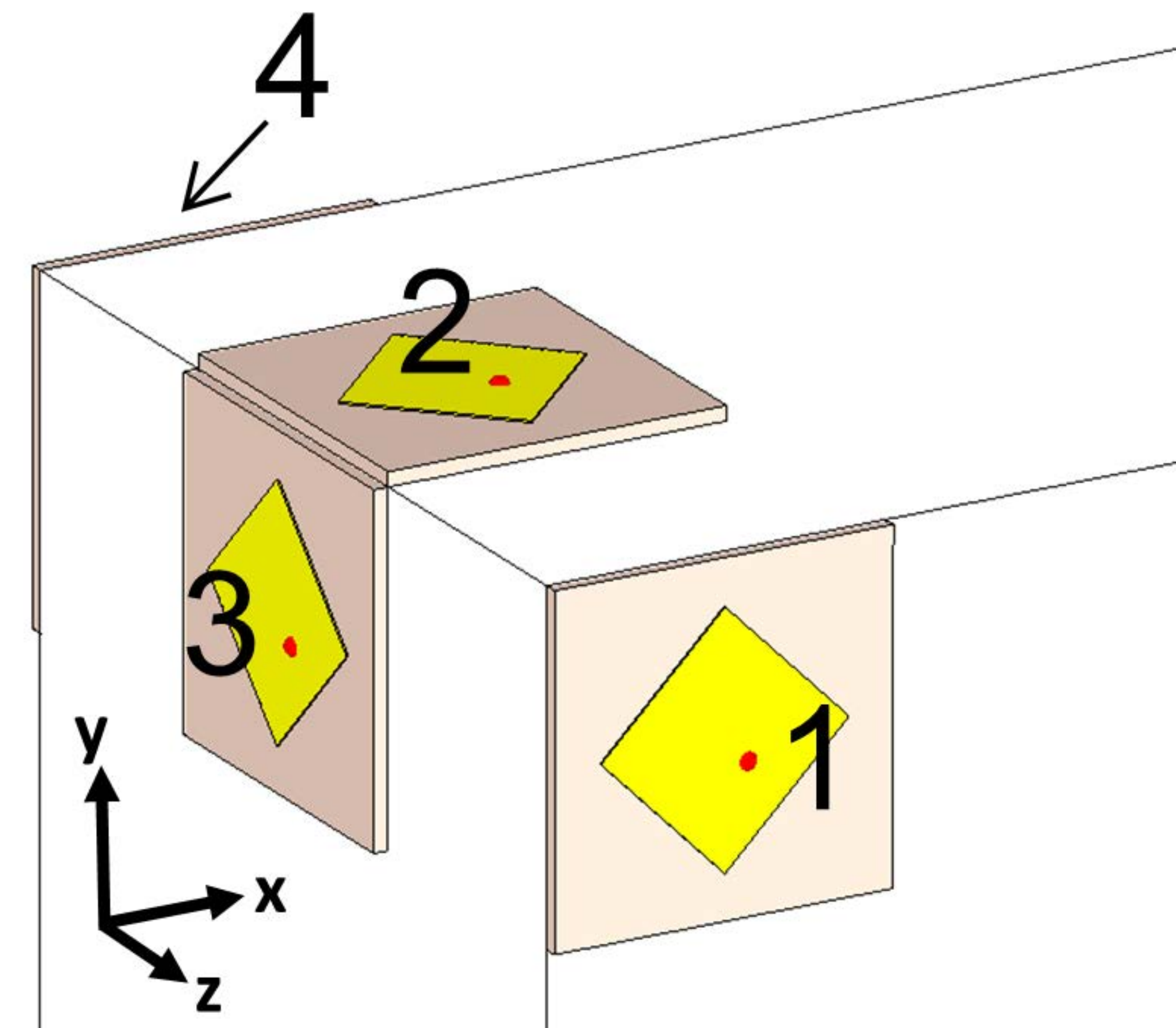}
        \label{fig:antenna_configuration_DA_left}}
   \subfigure[]{\includegraphics[scale=0.12]{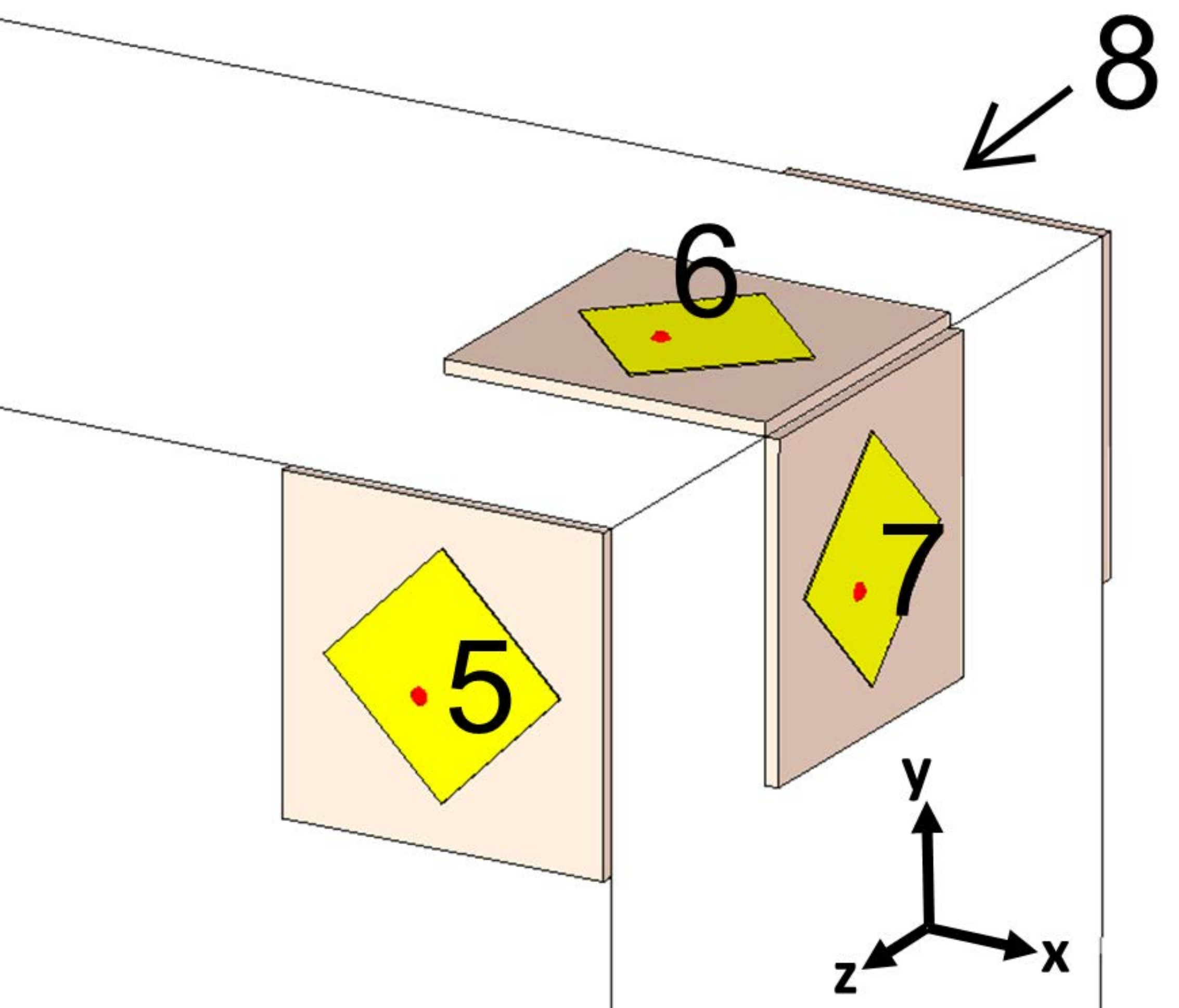}
        \label{fig:antenna_configuration_DA_right}}
  \caption{Arrangement of the antenna arrays on a mobile phone chassis with dimensions of $150 \times 75 \times 8~{\rm mm}^3$ in length, width and thickness: (a) ULA, (b) and (c) DA. The ULA radiates slanted polarizations to the ground, when the mobile phone is at this standing position.}
  \label{fig:antenna_configuration}
 \end{center}
\end{figure}

We consider two antenna arrays as practical examples of $60$~GHz arrays for mobile phones: 1) the ULA and 2) DA. Both are realized with square patch antennas oriented to radiate slanted polarizations when the mobile phone is at a standing position as shown in Fig.~\ref{fig:antenna_configuration}. The neighboring patch antennas of the ULA radiate the same polarizations to each other and are separated by half the free-space wavelength. The patch array is installed at the left-top corner of a mobile phone chassis as shown in Fig.~\ref{fig:antenna_configuration_ULA}. In the DA, the patch antennas are installed at each side of the two top corners of the chassis as illustrated in Figs.~\ref{fig:antenna_configuration_DA_left} and \ref{fig:antenna_configuration_DA_right}. The mobile phone has dimensions of $150 \times 75 \times 8~{\rm mm}^3$ in length, width and thickness and is considered to be a ground plane of the antennas. The patch antennas are simulated on a $0.127$~mm thick Rogers 5880 substrate with a relative permittivity of $\epsilon_{\rm r} = 2.2$ and a $17~{\rm \mu m}$-thick copper layer. The antennas have a broadside gain of $G_{\rm b}=8$~dB in the best case. The whole structure was simulated in CST Microwave Studio. The far-field radiation patterns of the antennas show that the maximum backlobe radiation is weaker than the main lobe gain by at least $20$~dB due to the mobile phone chassis serving as an electrically large ground plane. The DA has more uniform illumination of the entire solid angle since the broadsides of antenna elements point different directions. It is however harder for DA to leverage the array gain properly than ULA because DA elements do not illuminate space with similar gains. 

\subsection{Orientations of the Mobile Phone}
\begin{figure}[t]
	\begin{center}
		\includegraphics[scale=0.4]{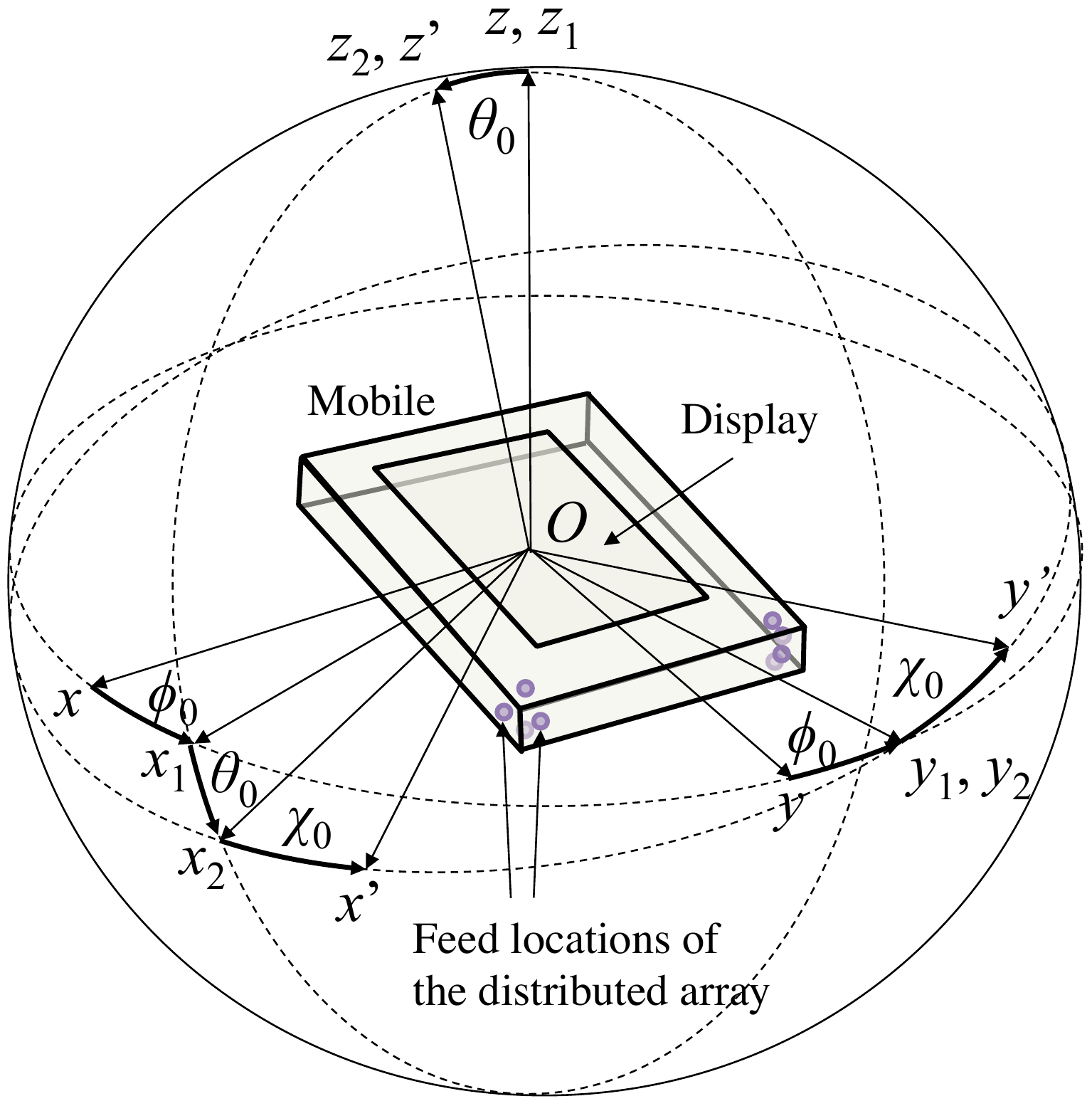}
		\caption{A coordinate system and three rotational axes defining the mobile phone orientation.}
		\label{fig:antenna_geometry}
	\end{center}
\end{figure}

Different orientations of the mobile phone are taken into account to analyze realistic operational scenarios. Figure~\ref{fig:antenna_geometry} shows the coordinate system and a base orientation of the mobile phone where the long-side of the mobile phone chassis is along the $y$-axis, while the display faces the $+z$-direction. Orientation of the mobile phone is determined by rotating the coordinate system through three angles, $\phi_0$, $\theta_0$ and $\chi_0$ in Fig.~\ref{fig:antenna_geometry}, while fixing the mobile phone. The three angles rotates the original coordinate system $(x,y,z)$ around $z$, $y_1$ and $z_2$ axes, respectively, so that the new coordinate system becomes $(x',y',z')$~\cite{Hansen98_book}, Appendix A2. The three angles $(\phi_0,\theta_0,\chi_0)$ are set such that the longitudinal axis of the mobile is along a line of every $45^\circ$ azimuth between $0^\circ$ and $315^\circ$ and of the polar angle at $0^\circ, 45^\circ$ and $90^\circ$. The radiation patterns of antenna elements on the rotated coordinate system, $\boldsymbol{E}_{\rm m}$, are derived from those of the base orientation on the original coordinate system as detailed in the Appendix of present paper. 

\subsection{Finger Effects}
We also consider cases when a finger covers one of the antenna elements. A finger covering an antenna gives rise to the reduction of the broadside gain of the patch antenna by 18 to 25~dB~\cite{Heino16_EuCAP}. 
A finger is modeled as a single layer with an elliptic cross section and is separated by $3$~mm from the antenna in our simulations to avoid severe reduction of radiation efficiency. When a fingertip points to one of the antennas, other antennas also suffer from shadowing due to the finger, but not as severely as the one covered with the finger. The worst input impedance matching at $60$~GHz among patch antennas is $9.2$~dB of return loss under the presence of a finger for ULA, while the same is $14.3$~dB for DA. The return losses are much greater without a finger. The worst isolation between patch antennas is $16.0$ and $35.3$~dB for ULA and DA, respectively, under the presence of a finger. The isolation is higher if there is no finger. The finger does not have much effects on the matching and isolation, as we ensured a minimum clearance of $3$~mm between a finger and patch antennas.

\section{Radio Propagation Simulations}
We introduce a channel sounding campaign and a ray-tracer in this section. The measured channels from the sounding campaign serve as a reference of experimental evidence that our ray-tracer aims at producing. Once the ray-tracer is qualified to produce realistic channels, it is used to generate multipath components (MPCs) at different mobile locations in the same site. Only a few technical details of the channel sounding are given here, followed by a calibration of our ray-tracer for the same site. The latter is the new contents that have not been published in the literature and hence the main focus of the present section.

\begin{figure}[tb]
	\centering 
	\includegraphics[scale=0.45]{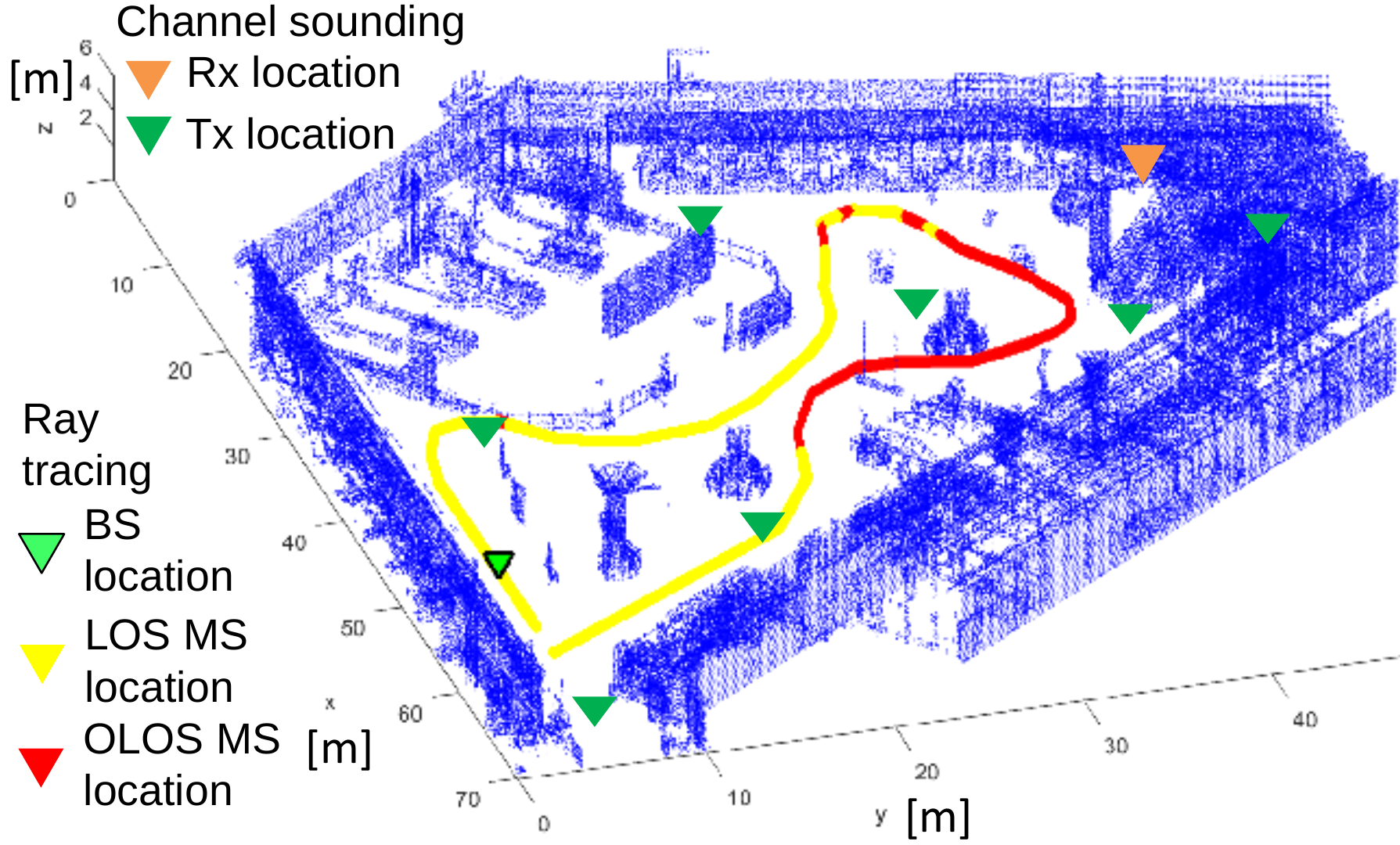}
	\caption{Floor plan of the small-cell site in an airport.}
	\label{fig:simulation_airport_floorplan}
\end{figure}

\subsection{Channel Sounding}
We chose a check-in hall of an airport as a representative small-cell scenario. A summary of multi-frequency channel sounder, and the channel sounding in the airport is given in~\cite{Vehmas16_VTCF}. A floor plan of the airport check-in hall where the channel sounding was performed is depicted in Fig.~\ref{fig:simulation_airport_floorplan}. To avoid human blockage effects, the measurements were conducted in the evening and at night. The transmit (Tx) antenna is placed at 12 different locations altogether though not all the Tx locations are mentioned in Fig.~\ref{fig:simulation_airport_floorplan}; while the receive (Rx) antenna was fixed near one of the walls overlooking the hall at an elevated floor. The measurements covered both the line-of-sight (LOS) and non-LOS (NLOS) channels, while the three-dimensional link distance varies from 18.8 to 107.2 m. The Tx and Rx antennas are $1.58$ and $5.68$~m high above the floor of the main hall. The Tx antenna is an omni-directional biconical antenna with $2$~dBi gain and elevation beamwidths of $60^\circ$, while the Rx antenna is a directive sectorial horn antenna with $19$~dBi gain and $10^\circ$ and $40^\circ$ azimuth and elevation beamwidths, respectively. Both antennas radiate and receive the vertical polarization mainly. The broadside of the Rx antenna was scanned over the azimuth angle from $20^\circ$ to $160^\circ$ with $5^\circ$ steps, and at $0^\circ$ and $-20^\circ$ elevation angles. The sounding was made with $4$~GHz of bandwidth centered at $61$~GHz. Power delay profiles (PDPs) are synthesized from a set of wideband directional channel sounding~~\cite{Vehmas16_VTCF} for calibration of the ray-tracer detailed in the next subsection.

\subsection{Optimization of the Ray-Tracer}
Our in-house ray-tracer simulates multipath channels for a large number of links between BS and MS. The ray-tracer is based on accurate descriptions of the environment in the form of point clouds, obtained by laser scanning~\cite{Jarvelainen16_TAP}, and is capable of simulating relevant propagation mechanisms such as specular reflections, diffraction, diffuse scattering and shadowing. Specular reflections are first identified by finding points lying inside the Fresnel zone between the MS and image of the BS, and then checking if a normal vector of a local surface formed by a group of points supports the specular reflection. Once identified, the reflection coefficients are calculated using the Fresnel equations. Shadowing objects are similarly detected by searching for points within the Fresnel zone for a given path. The ray-tracer provides azimuth and polar angles of arriving MPCs at the MS as well as the co-polarized magnitude of path gain and propagation delay time, i.e., $\left\{ \phi_{l}, \theta_{l}, \alpha^{VV}_{l}, \tau_l \right\}_{l=0}^{L_{\rm p}}$, as outputs where $L_{\rm p}$ is a number of MPCs in a BS-MS link; the path index $l=0$ is allocated to a LOS path. In the present simulations, we take into account the LOS path as well as first and second order specular reflections. Diffuse scattering is found to be of minor effects in the present case~\cite{Vehmas16_VTCF}.

\begin{figure}[t]
 \begin{center}
   \subfigure[]{\includegraphics[scale=0.4]{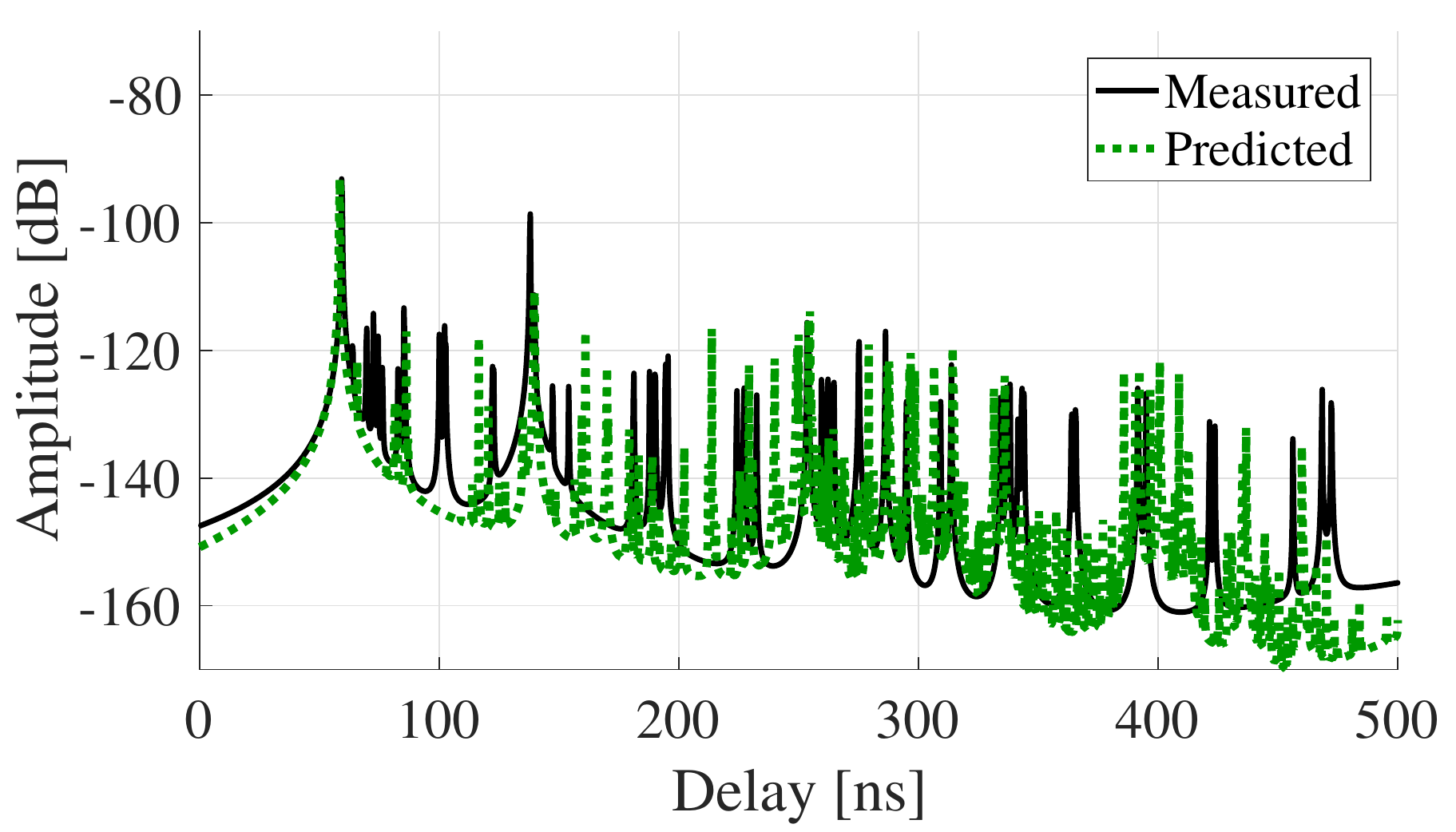}
        \label{fig:pdp_airport_meas_sim}}
   \subfigure[]{\includegraphics[scale=0.33]{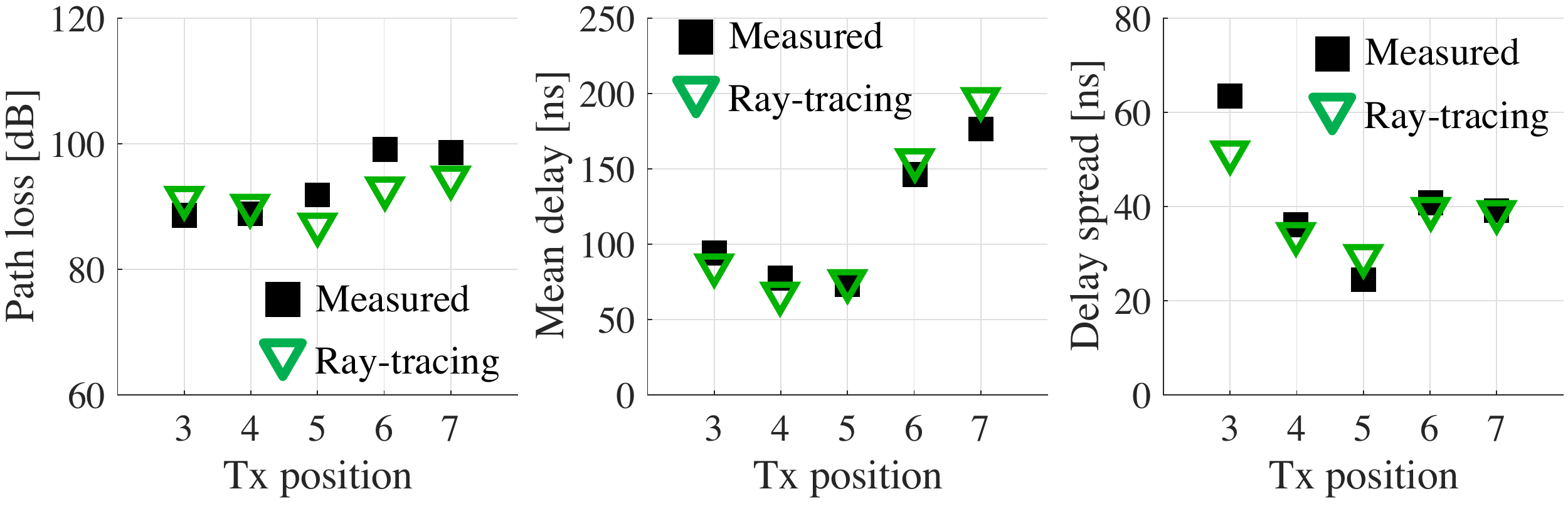}
        \label{fig:simulation_optimization_airport_60GHz}}
  \caption{(a) Measured and simulated PDP for Tx4-Rx link. (b) Large-scale parameters for measured and optimized simulated channels at 60 GHz.}
  \label{fig:meas_sim}
 \end{center}
\end{figure}

In order to ensure that the ray-tracer reproduces measured channels as accurate as possible, we set permittivity $\varepsilon_{\rm r}$ of objects in the environment so that the reproduced channels resemble the measured ones. Optimum $\varepsilon_{\rm r}$ is found by first calculating the path amplitude with different $\varepsilon_{\rm r}$ ranging from 2 to 6, then deriving the band-limited PDP from the paths and finally minimizing the difference between measured and simulated delay spreads. The shadowing attenuation loss $L_{\rm a}$ for small objects in the environment is chosen heuristically, resulting in $L_{\rm a}=20$~dB. Paths propagating through walls were assigned with very high attenuation losses because these paths do not contribute to the received power. Optimization yielded $\varepsilon_{\rm r}=3.6$, leading to agreement of the measured and simulated PDP shown in Fig.~\ref{fig:pdp_airport_meas_sim} for one of the measured links. A comparison between measured and simulated pathloss, mean delay and delay spreads is shown in Fig. \ref{fig:simulation_optimization_airport_60GHz}.% and Table \ref{tab:simulation_optimization_airport}. 

%\begin{table}[tb]
%	\centering
%	\caption{Average measured and simulated large-scale parameters.}
%	\label{tab:simulation_optimization_airport}
%	\begin{tabular}{ccccccc}
%		\toprule
%		\multirow{2}{*}{\bf Frequency} & \multicolumn{2}{c}{\bf Path loss [dB]} & \multicolumn{2}{c}{\bf Mean delay [ns]} & \multicolumn{2}{c}{\bf Delay spread [ns]} \\
%		& \bf meas. & \bf sim. & \bf meas. & \bf sim. & \bf meas. & \bf sim. \\
%		\midrule
%		60 GHz & 91.6 & 90.6 & 113.8 & 115.8 & 40.9 & 39.5 \\
%		\bottomrule
%	\end{tabular}
%\end{table} 

With the optimum parameters, MPCs are generated for links defined by BS and MS locations in Fig.~\ref{fig:simulation_airport_floorplan}. The BS is placed $1$~m from a wall at a height of $5.7$~m. The mobile is placed at a height of $1.5$~m at every $0.6$~m over a route. %, resulting in the distance between the base and mobile stations varying from $4$ to $64$ m. 
In total, we simulated 2639 links, including $1816$ LOS and $823$ obstructed LOS (OLOS). The polarimetric complex amplitudes of each MPC is generated statistically from $\alpha^{VV}$ estimates of the ray-tracer as $\alpha^{HH}= \alpha^{VV}$ and
\begin{equation}
 \alpha^{HV} = \alpha^{VH} = \alpha^{VV}/\mathrm{XPR},
\end{equation}
where $\alpha^{VH}$ for example denotes a complex amplitude of a horizontally-transmitted and vertically-received path; $\mathrm{XPR}$ is a cross-polarization ratio (XPR) of an MPC modeled from polarimetric channel sounding~\cite{Karttunen17_WCL} as
\begin{equation}
\mathrm{XPR}|_{\mathrm{dB}} \sim \mathcal{N}(\mu_2(L_\mathrm{ex}),\sigma_2^2),
\label{eq:XPR2}
\end{equation}
\begin{equation}
\mu_2(L_\mathrm{ex}) = \begin{cases} \alpha_2\cdot L_\mathrm{ex}+\beta_2, & \mbox{if } L_\mathrm{ex} \leq -\beta_2/\alpha_2 \\ 0, & \mbox{if } L_\mathrm{ex} > -\beta_2/\alpha_2 \end{cases},
\label{eq:mu2}
\end{equation}
where $\mu_2(L_\mathrm{ex})$ is the mean, $\sigma_2^2$ is the variance of the XPR model; $\alpha_2=-0.6, \beta_2=35$ and $\sigma_2=4$ were used~\cite{Karttunen17_WCL}. The excess loss $L_\mathrm{ex}$ of the MPC is defined as $L_\mathrm{ex}=|\alpha^{VV}|^2-\mathrm{FSPL}(\tau)$, where $\mathrm{FSPL}(\tau)$ is the free space path loss. For a LOS path $l=0$, ${\rm XPR}=\infty$.

\subsection{Human Torso Shadowing}
\begin{figure}[t]
 \begin{center}
   \subfigure[]{\includegraphics[scale=0.4]{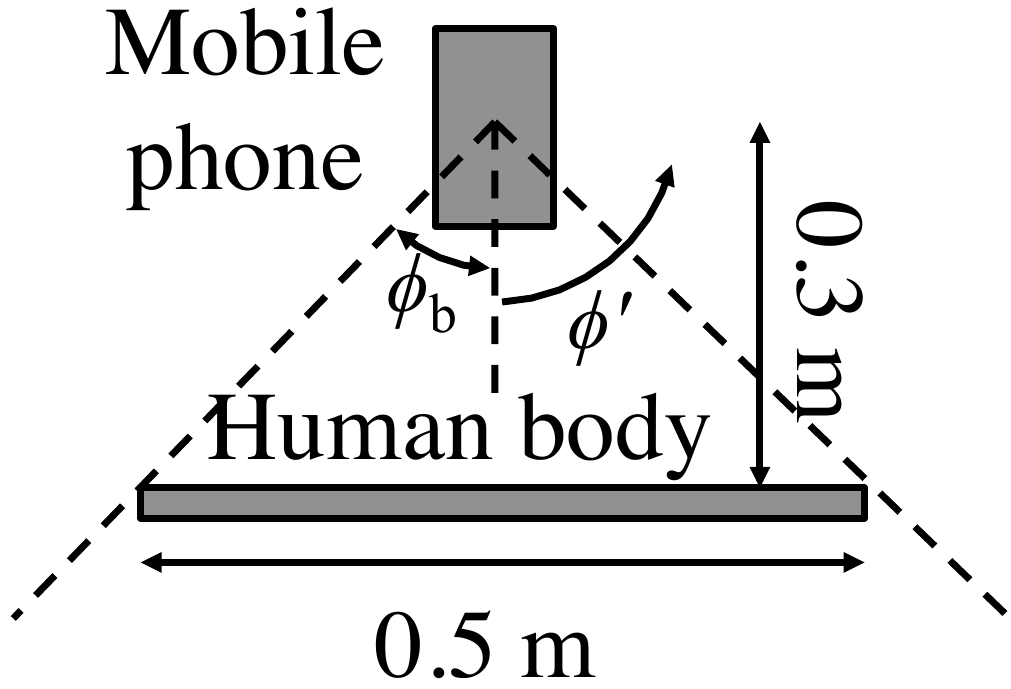}
        \label{fig:shadowing_geometry}}
   \subfigure[]{\includegraphics[scale=0.13]{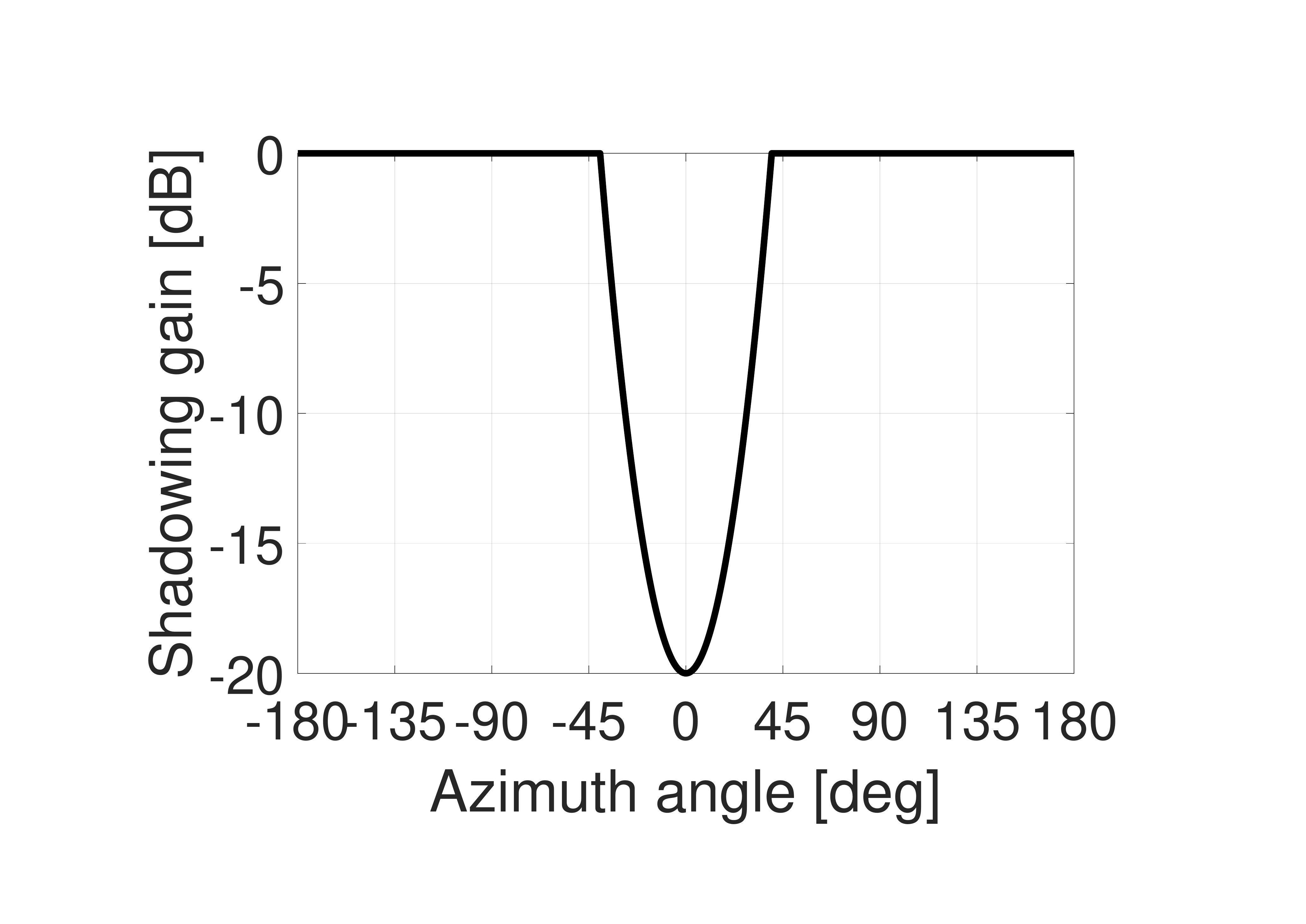}
        \label{fig:shadowing_loss}}
  \caption{(a) Geometry of the mobile phone and human torso. (b) Canonical model of shadowing losses at 60~GHz for a human torso.}
  \label{fig:shadowing}
 \end{center}
\end{figure}

At $60$~GHz, it is necessary to include a link blockage effect due to a human body holding the mobile phone. 
A simple canonical model of the link blockage due to a human body is used. 
A relative geometry of the human body to the mobile phone is defined in Fig.~\ref{fig:shadowing_geometry}, where the width of the human body and the separation between the body and mobile phone is $0.5$ and $0.3$~m, respectively.
The human blockage loss is defined by
\begin{equation}
 L_{\rm body} = \max \left( 0, L_{\rm b} \left\{ 1- \left( \frac{\phi-(\phi_0-\pi)}{\phi_{\rm b}} \right)^2 \right\} \right) {\rm dB},
\end{equation}
where $\phi$ is the azimuth angle of arrival of an MPC, $\phi_{\rm b} = 39.8^\circ$ is the azimuth angle of the body torso seen from the mobile phone as defined in Fig.~\ref{fig:shadowing_geometry}; $L_{\rm b} = 20$~dB is the maximum body shadowing loss. The model provides different losses depending on the azimuth angle of arrival of MPCs.

\section{Total Array Gain}
\subsection{Definition}
It is possible to define a total array gain of the antenna at MS locations after we have defined the polarimetric complex gain of the antenna radiation patterns, $\boldsymbol{E}_{\rm m}$, and parameters of MPCs, $\left\{ \phi_{l}, \theta_{l}, \boldsymbol{\alpha}_{l}, \tau_l \right\}_{l=1}^{L_{\rm p}}$, where $\boldsymbol{\alpha}_{l} \in \mathbb{C}^{2\times 2}$ is a polarimetric complex gain of an $l$-th MPC. Assuming downlink, the output signal $\boldsymbol{y}$ observed at a mobile antenna array is expressed as
\begin{equation}
 \boldsymbol{y} = \boldsymbol{h}x + \boldsymbol{n},
\end{equation}
where $x$ is an input voltage to a base station antenna, $\boldsymbol{n}, \boldsymbol{h} \in \mathbb{C}^N$ are vectors comprised of noise voltage observed at the antenna array and radio channel transfer functions, respectively, $1 \le n \le N$ is an index of an MS antenna. The $n$-th entry of $\boldsymbol h$ is given by
\begin{equation}
\label{eq:h}
 h_n = \sum_{l=0}^{L_{\rm p}} \boldsymbol{E}^{\rm H}_{{\rm m},n}(\phi_l, \theta_l) \boldsymbol{\alpha}_l \boldsymbol{E}_{\rm b} e^{\j \xi_l},
\end{equation}
where $\boldsymbol{E}_{{\rm m},n}$ is the polarimetric complex radiation pattern of the $n$-th antenna at a mobile, $\cdot^{\rm H}$ is Hermitian transpose. See Appendix for their definitions; $\boldsymbol{E}_{\rm b} = \left[ 1 ~ 1 \right]^{\rm T}/\sqrt{2}$ represents an ideal dual-polarized isotropic antenna at BS, $\xi$ is a uniformly distributed random phase over $[0~2\pi)$, which is set to $0$ for an LOS path;  $\cdot^{\rm T}$ denotes transpose. Adding the random phase leads to small-scale realizations of $\boldsymbol h$. We consider MRC assuming that a moving speed of the mobile is modest so that instantaneous channel information is available at MS, and that the link is noise-limited and hence the mobile aims at maximizing a signal-to-noise ratio. The combining weights are given by $\boldsymbol{w} = \boldsymbol{h}^H / || \boldsymbol{h} ||$, leading to the total array gain as,
\begin{equation}
G_{\rm a} = 10 \log_{10} \left( E_{\boldsymbol{h}} \left[ \left| \boldsymbol{h}\boldsymbol{w} \right|^2 \right] / P_{\rm o} \right) {\rm dB},
\label{eq:array_gain}
\end{equation}
where $E_{\boldsymbol{h}} [\cdot]$ is the Ensemble averaging over small-scale realizations of $\boldsymbol{h}$ and $P_{\rm o}$ is an omni-directional link gain denoted by
\begin{equation}
\label{eq:omni_pathloss}
P_{\rm o} = \sum^{L_{\rm p}}_{l=0} |\boldsymbol{\alpha}_l|^2,
\end{equation}
for a single mobile location. The total array gain includes averaged gains of all antenna elements in the array, as well as the gains attributed to signal precoding and combining. The total array gain is distinct from the conventional array gain in that its value is defined uniquely. The conventional array gain is practically not defined uniquely as it depends on the choice of a reference antenna element in an array. Powers from each antenna element in an array practically vary significantly depending on its polarization, orientation and element types. The total array gain allows fair comparison of phased antenna arrays consisting of different element types and configurations.

\begin{figure}[t]
 \begin{center}
   \subfigure[]{\includegraphics[scale=0.13]{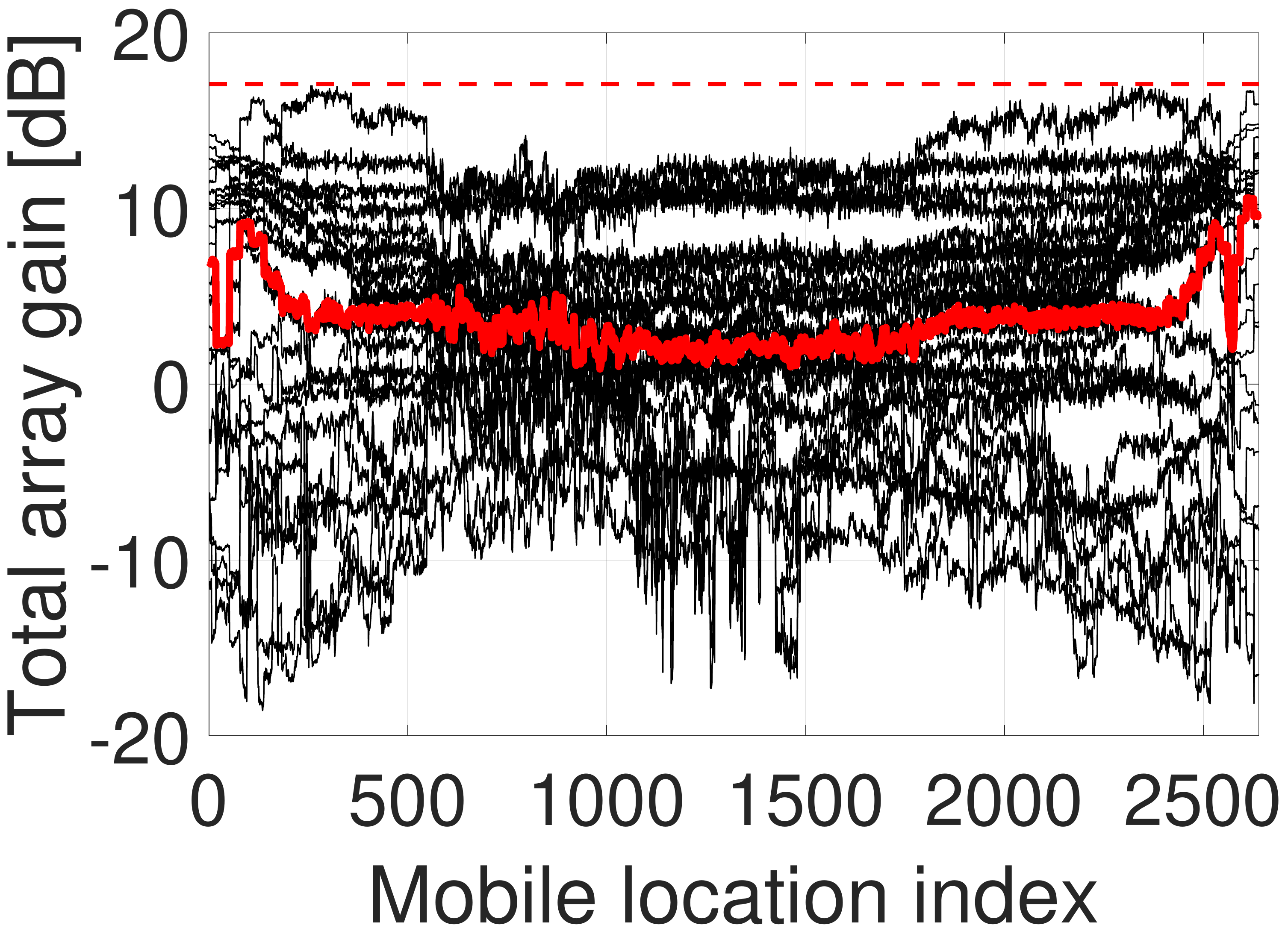}
        \label{fig:gain_unif_wo}}
   \subfigure[]{\includegraphics[scale=0.13]{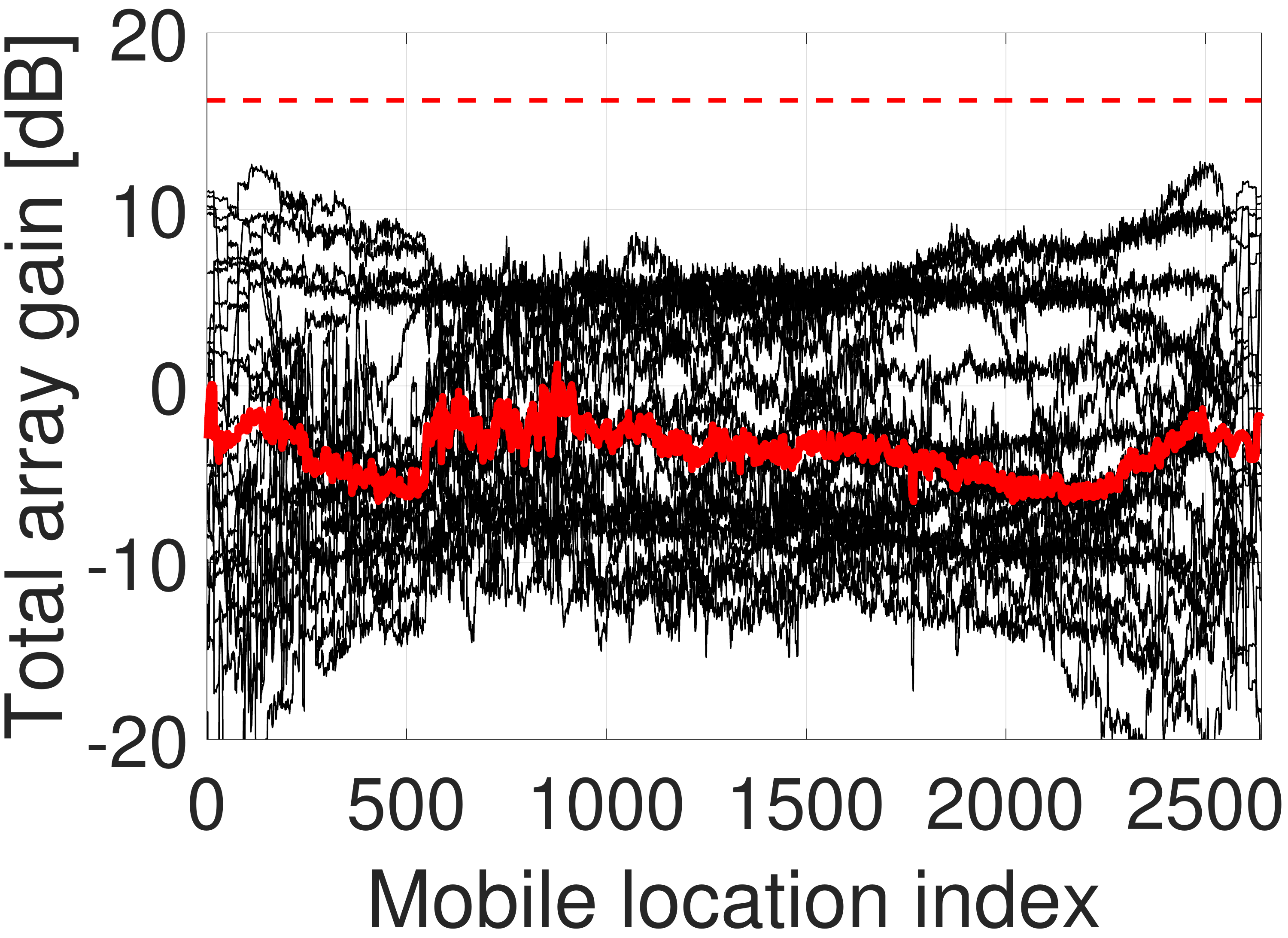}
        \label{fig:gain_unif_finger_body}}
   \subfigure[]{\includegraphics[scale=0.13]{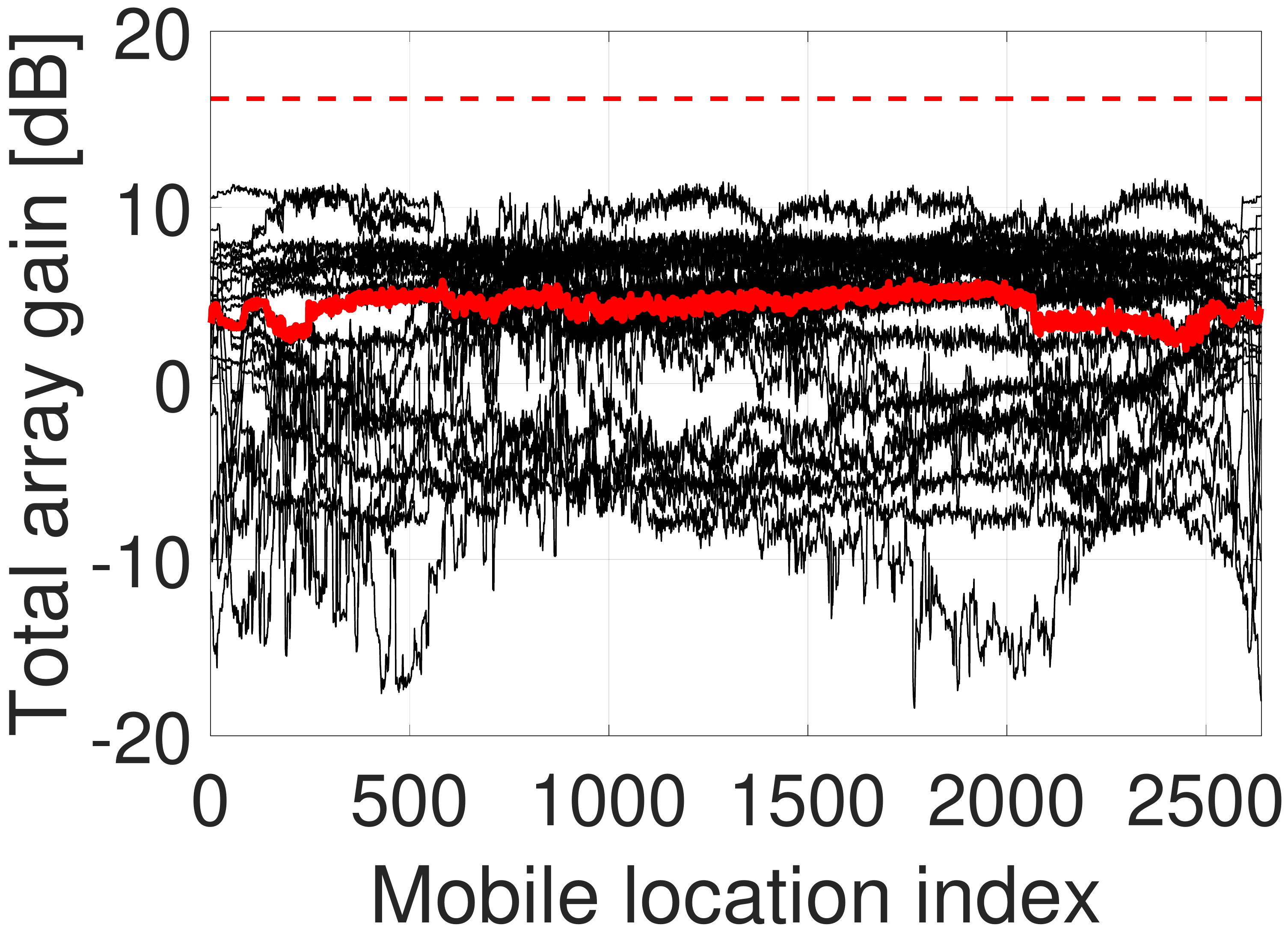}
        \label{fig:gain_dist_finger_body}}
    \subfigure[]{\includegraphics[scale=0.13]{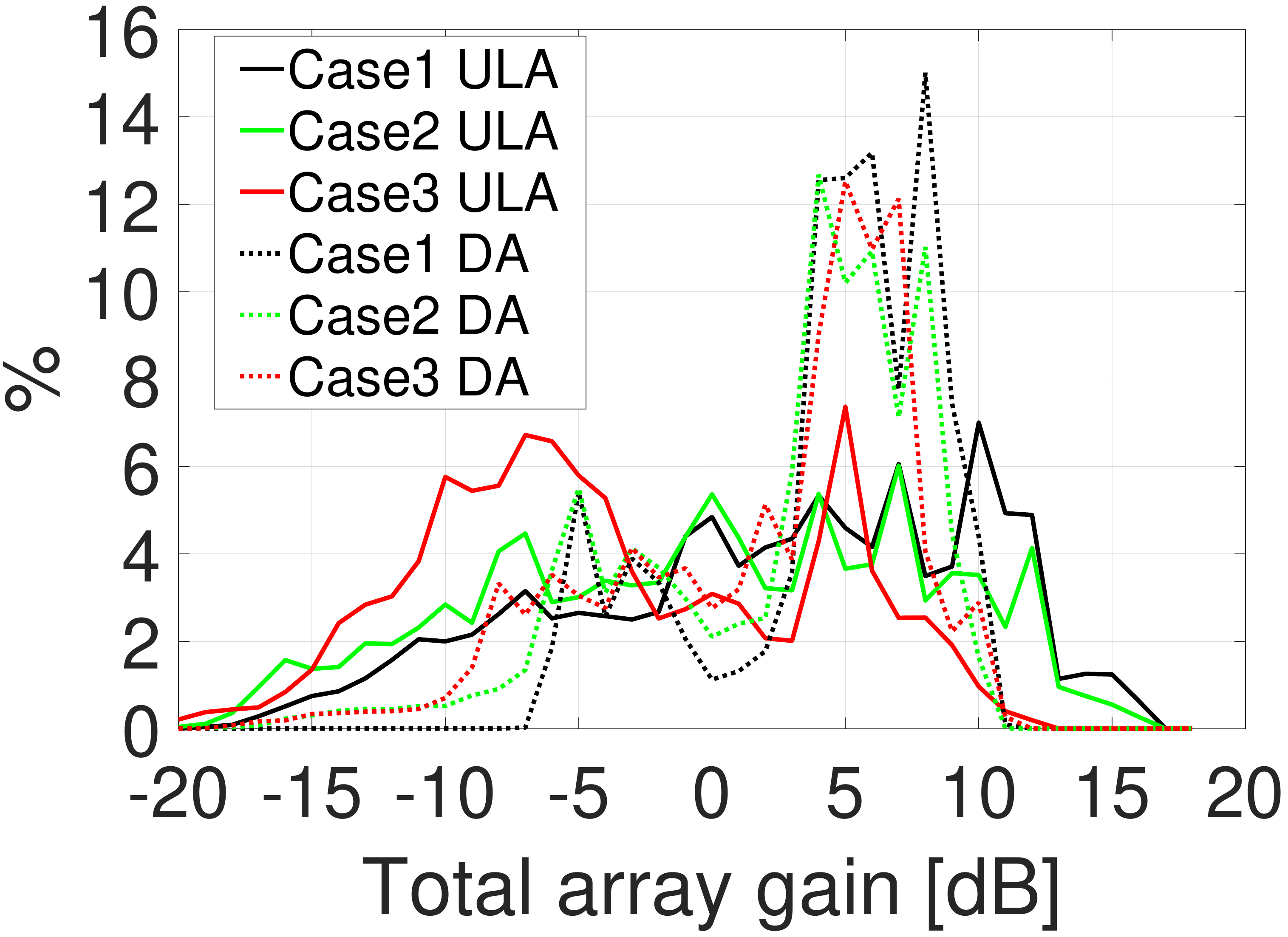}
 	\label{fig:histogram_array_gain}}
  \caption{(a) Total array gain of ULA in free space. (b) The same with human torso and finger shadowing and (c) total array gain of DA with human and finger shadowing. (d) Histogram of the total array gain. In (a)--(c), the red solid and dashed lines are median of the total array gain across different antenna orientations and the maximal gain of $8$-element patch antenna array. In (d), cases 1, 2 and 3 corresponds to the array in free space, with human body shadowing and with body and finger shadowing.}
  \label{fig:result_path_gain}
 \end{center}
\end{figure}
\begin{table}[t]
 \begin{center}
  {\caption{Statistics of the total array gain for ULA and DA across 24 different orientations and 2639 locations of the MS. Cases 1, 2 and 3 corresponds to the array in free space, with human body shadowing and with body and finger shadowing} 
  \label{table:MEG}}
   \begin{tabular}{c|ccc|ccc} \hline
    		Unit: dB	& \multicolumn{3}{c|}{Uniform linear array} 	& \multicolumn{3}{c}{Distributed array} \\
    				& Case 1 	& Case 2 	& Case 3	& Case 1 	& Case 2 	& Case 3 \\ \hline
		LOS &&&&&&\\		
    		peak	& $14.9$ 	& $12.9$	& $8.4$		& $10.3$	& $9.7$		& $10.2$ \\
			median	& $3.8$		& $0.8$ 	& $-4.1$	& $5.8$		& $4.6$ 	& $4.5$ \\
    		outage	& $-12.0$ 	& $-15.5$	& $-14.1$	& $-4.8$	& $-10.1$	& $-10.2$ \\ \hline
		OLOS &&&&&&\\		
			peak	& $11.8$	& $11.6$ 	& $6.3$		& $10.1$	& $9.3$ 	& $9.6$ \\
			median 	& $2.7$ 	& $1.5$ 	& $-2.7$	& $5.9$		& $5.0$ 	& $4.5$ \\ 
			outage	& $-8.1$	& $-12.4$ 	& $-12.0$	& $-5.1$	& $-6.7$ 	& $-7.4$ \\ \hline
     \end{tabular}
 \end{center}
 \end{table}

\subsection{Results and Discussions}
First, the total array gain of the ULA without human body and finger effects, called ``Case 1" hereinafter, is shown in Fig.~\ref{fig:gain_unif_wo} for 2639 MS locations of the small-cell site in the airport. The black lines correspond to total array gains at particular antenna orientations after MRC~\eqref{eq:array_gain}. The red solid and dashed lines correspond to the median gain and the maximum total array gain of an $8$-element patch array, $G_{\rm b}+10\log_{10} 8 = 17.2$~dB, respectively. The antenna orientation causes $\pm 15$~dB variation of the total array gain around the median. 

The total array gain of ULA is illustrated in Fig.~\ref{fig:gain_unif_finger_body} when the human body and finger shadowing is considered; we call it ``Case 3" in the following. Table~\ref{table:MEG} summarizes the total array gain at peak, median and outage levels for the ULA with three cases. The ``Case 2" in the table correspond to links with body shadowing. The table shows that the median gain drops by about $8$~dB due to body and finger compared to the free space case. The peak and outage levels are defined by largest and smallest $2$~\% total array gains. It is worth noting that the outage total array gain improves for ULA when finger shadowing is present, because the finger guides the radiated energy to directions where it is otherwise not possible to reach. Diffraction on finger surface mainly produces such radiation~\cite{Heino16_EuCAP}.

The gain of DA with human body and finger shadowing effects is shown in Fig.~\ref{fig:gain_dist_finger_body}. The right-most column of Table~\ref{table:MEG} shows the corresponding gain statistics. The results show better median gain of DA than ULA by about $8$~dB when both the body and finger are present. Even though the peak gain of the DA is smaller than that of the ULA, the median and outage gain of DA is up to $8$~dB higher than that of the ULA, showing robustness of DA for capturing energy delivered by MPCs across different orientations of the mobile phone. 

Finally, Fig.~\ref{fig:histogram_array_gain} shows a histogram of total array gain across different MS locations and orientations. The plot shows that gains of DA concentrate on the positive side, while those of ULA are more scattered across positive and negative sides.

It is worth pointing out that total array gains here assume idealistic lossless hardware for analog beamforming; practical beamforming with analog phase shifting networks, regardless of passive or active, reduces a signal-to-noise ratio and hence the total array gains of practical beamforming hardware will be smaller than the values presented in this paper.

\section{Concluding Remarks}
The present paper quantified gains of mobile phone antenna arrays at $60$~GHz equipped with analog beamforming and a single baseband unit. The total array gain is defined by a received power at the array in excess to that of a single-element isotropic antenna. The total array gain circumvents the ambiguity of conventional array gains depending on the choice of a reference antenna element in an array. Our analysis revealed that the total array gain is higher for $8$-element DA than the ULA by up to $8$ and $6$~dB in the median and outage levels, respectively. The ULA achieves the maximum total array gain when there is an LOS to a BS and no human shadowing is involved, while there are many postures of the mobile phone and hence the antenna array where ULA cannot see the LOS, leading to a low gain of $-15$~dB. On the other hand, DA can always illuminate the entire solid angle, making it possible to keep the median and outage noticeably higher than ULA. The present study therefore shows robustness of DA as an mm-wave antenna array on a mobile device. In our forthcoming works, we will address the total array gain when the BS also performs beamforming. 

\section*{Appendix: Rotation of a Mobile Phone}
When the mobile phone is at the base orientation shown in Fig.~\ref{fig:antenna_geometry}, we call the corresponding radiation patterns of antenna elements as base patterns. Denoting the base patterns as
$\boldsymbol{E}_{\rm o} = \left[ \boldsymbol{e}_{V}~\boldsymbol{e}_{H}\right]^T$ where $\boldsymbol{e}_{\alpha}$, $\alpha = V~{\rm or}~H$ denotes the far-field radiation pattern of the vertical or horizontal polarizations. They refer to the electric field components tangential to the polar and azimuth angles of the antenna coordinate system, respectively. The radiation pattern vectors are defined by 
\begin{equation}
 \boldsymbol{e}_{\alpha} = \left[ E_{\alpha,\boldsymbol{\Gamma}_1}  \cdots  E_{\alpha,\boldsymbol{\Gamma}_l}  \cdots E_{\alpha,\boldsymbol{\Gamma}_L} \right]^T \in \mathbb{C}^{L},
 \end{equation}
where $\boldsymbol{\Gamma}_l = \left[ \theta_l ~ \phi_l \right]$ refers to the $l$-th pointing direction of the radiation pattern from the origin of the mobile phone, $1 \le l \le L$; $E_{\alpha,\boldsymbol{\Gamma}}$ denotes a complex gain for the respective polarization and pointing direction. The base pattern can be decomposed into a series of spherical harmonics coefficients as
\begin{equation}
 \label{eq:decomposition}
 {\boldsymbol E}_{\rm o} = \frac{k}{\sqrt{\eta}} \boldsymbol{Fq},
\end{equation}
where $k$ is wavenumber, $\eta$ is a wave impedance of vacuum, ${\boldsymbol F} \in \mathbb{C}^{2L \times J}$ is a matrix denoting $\phi$ and $\theta$ fields of each spherical harmonic as
\begin{equation}
 \label{eq:f}
 \boldsymbol{F} =
 \begin{pmatrix}
  \boldsymbol{f}_{V,1-11} &  \boldsymbol{f}_{V,2-11}   & \cdots &  
  \boldsymbol{f}_{V,1mn} & \boldsymbol{f}_{V,2mn} & \cdots \\
  \boldsymbol{f}_{H,1-11} &  \boldsymbol{f}_{H,2-11}   & \cdots &  
  \boldsymbol{f}_{H,1mn} & \boldsymbol{f}_{H,2mn} & \cdots \\
  \end{pmatrix},
\end{equation}
where $-n \le m \le n$ and $1 \le n \le N$; $m$ and $n$ are the spherical wavemode indices and $N$ is the total number of considered $n$-modes.  The total number of the spherical wavemodes amounts to $J=2N(N+2)$, \cite{Hansen98_book}, pp.15.  In~\eqref{eq:f},
\begin{equation}
 \boldsymbol{f}_{\alpha,smn} = \left[ f_{\alpha,\boldsymbol{\Gamma}_1,smn} \cdots f_{\alpha,\boldsymbol{\Gamma}_l,smn} \cdots f_{\alpha,\boldsymbol{\Gamma}_L,smn} \right]^T \in \mathbb{C}^{L},
\end{equation}
where $s=1$ or $2$.  Furthermore, under the $e^{\j \omega t}$ time convention where $\omega$ is an angular frequency of a radio frequency signal,\footnote{The time convention in~\cite{Hansen98_book} is $e^{-\j \omega t}$.}
\begin{eqnarray}
\label{eq:f1}
 f_{V,\boldsymbol{\Gamma},1mn}  & = & k_{mn} (-\j)^{n+1} 
 \left( \frac{\j m\bar{P}_n^{|m|}(\cos{\theta}) }{\sin{\theta}} \right), \\
 f_{H,\boldsymbol{\Gamma},1mn}  & = & k_{mn} (-\j)^{n+1} 
 \left( -\frac{\d }{\d \theta} \bar{P}_n^{|m|}(\cos{\theta}) \right), \\
 f_{V,\boldsymbol{\Gamma},2mn}  & = & k_{mn} (-\j)^n
 \left( \frac{\d}{\d \theta} \bar{P}_n^{|m|}(\cos{\theta}) \right), \\
 \label{eq:f4}
 f_{H,\boldsymbol{\Gamma},2mn}  & = & k_{mn} (-\j)^n
 \left( \frac{\j m\bar{P}_n^{|m|}(\cos{\theta}) }{\sin{\theta}} \right), \\
 k_{mn} & = & \sqrt{\frac{2}{n(n+1)}} \left( -\frac{m}{|m|} \right)^m e^{- \j m\phi} ,
 \end{eqnarray}
where $\bar{P}_n^{|m|}(\cdot)$ is the normalized associated Legendre function with the order of $m$.
Finally, in~\eqref{eq:decomposition}, ${\boldsymbol q} \in \mathbb{C}^{J}$ is a vector comprised of spherical harmonics coefficients. The vector can be solved in a least-squares manner using a pseudo-inverse of a matrix $\boldsymbol F$.

Now we consider rotation of the spherical harmonics~\eqref{eq:f1}-\eqref{eq:f4} through the Euler angles $\phi_0$, $\theta_0$ and $\chi_0$ that transform the original one $(x,y,z)$ into another coordinate system $(x', y', z')$ as defined in Fig.~\ref{fig:antenna_geometry}. The three angles are applied to rotate the original coordinate system around $z$, $y_1$ and $z_2$ axes, respectively in this order~\cite{Hansen98_book}, Appendix A2. The spherical harmonics of the rotated coordinate system is given by
\begin{equation}
f'_{\alpha, \boldsymbol{\Gamma}, smn} = \sum_{\mu = -n}^{n} e^{\j m \phi_0} d^n_{\mu m}(\theta_0) e^{\j \mu \chi_0} f_{\alpha, \boldsymbol{\Gamma}, s\mu n},
\end{equation}
where
\begin{eqnarray}
\nonumber
 d^n_{\mu m}(\theta) & = & \left\{ \frac{(n+\mu)!(n-\mu)!}{(n+m)!(n-m)!} \right\}^{\frac{1}{2}} \left( \cos{\frac{\theta}{2}} \right)^{\mu +m} \times \\
 &  &  \left( \sin{\frac{\theta}{2}} \right)^{\mu -m} P_{n-\mu}^{(\mu-m, \mu+m)}(\cos\theta),
\end{eqnarray}
where $P_{n-\mu}^{(\mu-m, \mu+m)}(\cos\theta)$ is the Jacobi polynomial~\cite{Hansen98_book}.

Finally, the radiation patterns of the antenna elements on a rotated coordinate system are derived using the same spherical harmonics coefficients as
\begin{equation}
 \label{eq:decomposition2}
 {\boldsymbol E}_{\rm m} = \frac{k}{\sqrt{\eta}} \boldsymbol{F}'\boldsymbol{q},
\end{equation}
where $\boldsymbol{F}'$ is given in a similar manner as~\eqref{eq:f} with the spherical harmonics on the rotated coordinate system.

\section*{Acknowledgement}
The authors would like to acknowledge the financial support from the Academy of Finland research project ``Massive MIMO: Advanced Antennas, Systems and Signal Processing at mm-Waves (M3MIMO)" (decision \#288670), as well as the support from the Nokia Bell Labs, Finland.

\bibliographystyle{IEEEtran}% bib style
\bibliography{ref}% your bib database

% that's all folks
\end{document}